\documentclass{pasj02} 
\usepackage[switch,mathlines]{lineno} 
\usepackage{amsmath}
\usepackage{url}

\newcommand{\cgsflux}{erg~s$^{-1}$~cm$^{-2}$}
\newcommand{\cgslumin}{erg~s$^{-1}$}
\newcommand{\phiorb}{\phi_{\textrm{orb}}}

\renewcommand{\thefootnote}{\alph{footnote}}
\DeclareRobustCommand{\ion}[2]{%
\relax\ifmmode
\ifx\testbx\f@series
{\mathbf{#1\,\mathsc{#2}}}\else
{\mathrm{#1\,\mathsc{#2}}}\fi
\else\textup{#1\,{\mdseries\textsc{#2}}}%
\fi}

\jyear{2026}
\Received{}
\Accepted{}

\begin{document} 

\title{Energy shift of Fe-K fluorescence lines due to low ionization demonstrated with XRISM in Centaurus X-3}

\author{Yutaro~\textsc{Nagai}
\altaffilmark{1}\altemailmark
\orcid{0009-0003-9261-2740}
\email{nagai.yuutarou.25r@st.kyoto-u.ac.jp} 
}
\author{Teruaki~\textsc{Enoto}
\altaffilmark{1,2}\altemailmark
\orcid{0000-0003-1244-3100}
\email{enoto.teruaki.2w@kyoto-u.ac.jp}
}
\author{Masahiro~\textsc{Tsujimoto}
\altaffilmark{3,4}
\orcid{0000-0002-9184-5556} 
}
\author{Hiroya~\textsc{Yamaguchi}
\altaffilmark{3,4}
\orcid{0000-0002-5092-6085}
}
\author{Yuto~\textsc{Mochizuki}
\altaffilmark{3,4}
\orcid{0000-0003-3224-1821}
}
\author{Ehud~\textsc{Behar}
\altaffilmark{5}
\orcid{0000-0001-9735-4873}
}

\author{Lia~\textsc{Corrales}
\altaffilmark{6}
\orcid{0000-0002-5466-3817}
}
\author{Paul~A.~\textsc{Draghis}
\altaffilmark{7}
\orcid{0000-0002-2218-2306}
}
\author{Ken~\textsc{Ebisawa}
\altaffilmark{3}
\orcid{0000-0002-5352-7178}
}
\author{Natalie~\textsc{Hell}
\altaffilmark{8}
\orcid{0000-0003-3057-1536}
}
\author{Timothy~R.~\textsc{Kallman}
\altaffilmark{9}
\orcid{0000-0002-5779-6906}
}
\author{Richard~L.~\textsc{Kelley}
\altaffilmark{9}
\orcid{0009-0007-2283-3336}
}

\author{Pragati~\textsc{Pradhan}
\altaffilmark{10}
\orcid{0000-0002-1131-3059}
}
\author{Shinya~\textsc{Yamada}
\altaffilmark{11}
\orcid{0000-0003-4808-893X}
}
\author{Toshiyuki~\textsc{Azuma}
\altaffilmark{12}
\orcid{0000-0002-6416-1212}
}
\author{Xiao-Min~\textsc{Tong}
\altaffilmark{13}
\orcid{0000-0003-4898-3491}
}

\altaffiltext{1}{Department of Physics, Graduate School of Science, Kyoto University, Sakyo-ku, Kyoto, Kyoto 606-8502, Japan}
\altaffiltext{2}{Center for Advanced Photonics, RIKEN, Wako, Saitama 351-0198, Japan}
\altaffiltext{3}{Institute of Space and Astronautical Science, Japan Aerospace Exploration Agency, Chuo-ku, Sagamihara, Kanagawa 252-5210, Japan}
\altaffiltext{4}{Department of Astronomy, Graduate School of Science, The University of Tokyo, Bunkyo-ku, Tokyo 113-0033, Japan}
\altaffiltext{5}{Physics Department, Technion, Haifa 32000, Israel}
\altaffiltext{6}{Department of Astronomy, University of Michigan, Ann Arbor, MI 48109, USA}
\altaffiltext{7}{MIT Kavli Institute for Astrophysics and Space Research, Massachusetts Institute of Technology, 70 Vassar St, Cambridge, MA 02139, USA}
\altaffiltext{8}{Lawrence Livermore National Laboratory, Livermore, CA 94550, USA}
\altaffiltext{9}{NASA's Goddard Space Flight Center, Greenbelt, MD 20771, USA}
\altaffiltext{10}{Department of Physics, Embry-Riddle Aeronautical University, Prescott, AZ 86301, USA}
\altaffiltext{11}{Department of Physics, Rikkyo University, Toshima-ku, Tokyo 171-8501, Japan}
\altaffiltext{12}{Atomic, Molecular and Optical Physics Laboratory, RIKEN, Wako, Saitama 351-0198, Japan}
\altaffiltext{13}{Center for Computational Sciences, University of Tsukuba, Tsukuba, Ibaraki 305-8573, Japan}

\maketitle

\begin{abstract}
 The Fe K$\alpha$ fluorescence line at 6.4~keV is a powerful probe of cold matter
 surrounding X-ray sources and has been widely used in various astrophysical contexts.
 The X-ray microcalorimeter spectrometer onboard XRISM can measure line shifts with
 unprecedented precision of $\sim$0.2~eV, equivalent to a line-of-sight velocity of
 $\sim$10~km~s$^{-1}$. At this level of accuracy, however, several factors that
 influence the line energy must be carefully considered prior to astrophysical
 interpretation. One such important factor is the ionization degree, Fe$^{q+}$.  The
 K$\alpha$ line shifts redward by $\sim$4~eV as $q$ increases from 0 (neutral) to 8
 (Ar-like). Additionally, the accompanying Fe K$\beta$ line at 7.06~keV shifts blueward
 by $\sim$30~eV from $q=0$ to 8. We demonstrate that this effect is actually observable
 in the XRISM data of the high-mass X-ray binary Centaurus X-3 (Cen X-3). We advocate
 that the differential energy shift between the K$\alpha$ and K$\beta$ line provides a
 robust estimate of $q$ by decoupling from other effects that shift the two lines in the
 same direction. We derived $q \sim 5$ (Sc-like) for the fluorescing matter by comparing
 the observation with atomic structure calculations of our own and in the literature. By
 accounting for the derived charge state and the corresponding shift in the rest-frame
 line energy, we made corrections for this effect and reached a consistent residual
 shift among the K$\alpha$, K$\beta$, and the optical measurement attributable to the
 systemic velocity of the system. Consequently, we obtained a new constraint on the
 location of the cold matter. This ionization effect needs to be assessed in all use
 cases of the Fe K$\alpha$ line shift beyond Cen X-3, and the proposed metric is
 generally applicable to all of them.
\end{abstract}

\section{Introduction}\label{s1}
The Fe K$\alpha$ fluorescent emission line at 6.4~keV is one of the most widely used
diagnostics in X-ray spectroscopy. The emission is produced by radiative decay
following the inner-shell ionization of neutral to low-ionized Fe$^{q+}$ (typically
below Ar-like with an ionization degree $q \le 8$, where $q=0$ is for neutral) in 
cold matter. The ionization source can be photons or charged particles above the
threshold energy of 7.11~keV. The line carries information about the ionizing source,
the cold matter, and the relation between them through its observed properties: i.e.,
the intensity or the equivalent width (EW), the energy shift, and the line
profile. Due to the rich cosmic abundance of Fe, the large fluorescence yield (0.355; \cite{Firestone1996})
, and the sufficiently high energy to penetrate through the interstellar
extinction of $N_{\rm H}$ $\sim 10^{22}$~cm$^{-2}$, the line has profound utility in almost all
fields of X-ray astronomy \citep{Kaastra1999}.

The utility is further enhanced by the advent of the X-ray microcalorimeter spectrometer
onboard the X-Ray Imaging and Spectroscopy Mission (XRISM; \cite{Tashiro2025}). With the
superb spectroscopic performance of the \textit{Resolve} instrument
\citep{Kelley2025,Ishisaki2025}, the Fe K$\alpha$ line is used in ways that were never
possible with previous CCD and grating spectrometers. The two intense lines of the
complex (K$\alpha_1$ and K$\alpha_2$ respectively at 6.404 and 6.391~keV for $q=0$) were
resolved \citep{Miller2025, Bogensberger2025}; the weak satellite lines (K$\alpha_{3,4}$
at 6.435~keV) were detected (Vander Meulen et al. in preparation); the bulk velocity of
$<$100~km~s$^{-1}$ was derived \citep{Chakraborty2025}; the radial velocity (RV)
curve of a wide binary was obtained (Naz\'{e} et al. submitted); Doppler tomography was
applied (Sameshima et al. submitted); the multiple components of different broadening
were separated \citep{Kammoun2025,XRISM2024}; and the timing of the line photons
relative to the continuum photons was constrained \citep{Mochizuki2024}.

These novel applications remind us that we need to know the intrinsic properties of the
line to the accuracy that the microcalorimeter spectra can achieve. Such caution is most
evident in the result of the \textit{Resolve} observation of the X-ray binary Centaurus
X-3 (Cen X-3). \citet{Mochizuki2024} revealed that the line energy is modulated by the
binary orbital motion. The RV curve of the line is consistent with the NS motion in
phase but with a slightly smaller amplitude than expected. However, the most striking
result is the measured RV offset of $-140 \pm 4$~km~s$^{-1}$, which was very tightly
constrained by observing the RV curve over the entire phase. The magnitude of the shift
(equivalent to 470~ppm at 6.4~keV) is significantly larger than all conceivable sources
of systematics, including the systemic velocity of the system derived from optical
spectra of the O star (130~ppm; \cite{Hutchings1979}), the energy determination accuracy
of \textit{Resolve} (30~ppm; \cite{Porter2025,Eckart2025}), the Earth's orbital motion around the
Sun (51~ppm), and the Sun's motion with respect to the local standard of rest (17~ppm)
at the time of the observation.

It is obvious that we need to revisit the assumptions regarding the intrinsic properties
of the line based on atomic physics. In particular, the assumption that the line is from
neutral ($q=0$) Fe can lead to incorrect conclusions, as the line energy can shift by
$\sim$4~eV (625~ppm) when Fe is ionized to $q = 8$ \citep{Palmeri2003}. The purpose of
this paper is to demonstrate, for the first time, that this effect actually occurs using
the same Cen X-3 data with \textit{Resolve}. A robust assessment is achieved by
combining it with the accompanying K$\beta$ fluorescence line at 7.06~keV, which became
accessible with \textit{Resolve} without contamination from the K edge feature at
7.11~keV. As we will show in this paper, the differential energy shift between the
K$\alpha$ and K$\beta$ lines behaves distinctly differently from other sources of
systematics. By taking this effect into account properly, we further show that the
systemic velocity of the system can be better reconciled among different results, and
the average $q$ value can be derived, which provides new constraints on the fluorescing
matter and its ionizing source.

\medskip

The outline of this paper is as follows. We start with a brief description of the
target, the instrument, the observation, and the data reduction in
section~\ref{s2}. A more detailed description can be found in the preceding paper
\citep{Mochizuki2024}. We present the results of the data analysis in section~\ref{s3}, in
which we characterize the observed K$\alpha$ and K$\beta$ lines. In section~\ref{s4}, we
evaluate the atomic effects for the line energy shifts (subsection~\ref{s4-1}). We further
discuss new constraints on fluorescing matter in the Cen X-3 system
(subsection~\ref{s4-2}). 
We summarize our findings and conclude with their astrophysical implications in section~\ref{s5}.

All errors throughout the paper are given for the 1$\sigma$ statistical uncertainty. We
take RV redshifts (lower shift in energy) as negative values to align with the preceding
work \citep{Mochizuki2024}, although this is contrary to astronomical conventions.

\section{Data}\label{s2}
Cen X-3 is an eclipsing high-mass X-ray binary with an orbital period of 2.09~days.  It
is composed of a neutron star (NS) and an O6--8 III giant V779~Cen
\citep{Krzeminski1974,Hutchings1979}. The XRISM observation was performed to cover an
entire orbital phase from February 13 to 16, 2024. The dates were chosen so that the
target could be observed with minimum interruption, with no Earth occultation seen from
XRISM orbiting the Earth in a precessing plane. As a consequence, the line-of-sight
velocity of XRISM around the Earth was very small at 0.5~km~s$^{-1}$ (1.7~ppm) in
peak-to-peak. The basic properties of the system and the observation are tabulated in
table~\ref{tab:par_summary}. Compared to other X-ray observations in the past
\citep{Naik2011,Sanjurjo2021,Tamba2025}, the X-ray flux remained relatively stable over
an entire orbit outside of eclipses, making it easy to interpret the observed RV curve.

We use the \textit{Resolve} instrument aboard XRISM \citep{Kelley2025,Ishisaki2025}. It
hosts an array of 6$\times$6 X-ray microcalorimeter pixels (pixels 0--35) placed at the
focal plane of the X-ray mirror assembly. One of the pixels (pixel 12) is displaced from
the array for calibration purposes. When all the array pixels are combined, it has a
spectral resolution of $R\equiv E/\Delta E=1300$ \citep{Porter2025} and an effective
area of 180~cm$^2$ \citep{Hayashi2024} at 5.9~keV, a band-pass coverage of 1.7--12~keV,
a very low background equivalent to $\lesssim$0.5 events per 5~eV for a 100~ks exposure
\citep{Mochizuki2025a}, and a $3^{\prime} \times 3^{\prime}$ field of view with a point
spread function of 1\farcm3. The spectroscopic performance surpasses those of any
previous spectrometers at the Fe K band, but the true uniqueness of \textit{Resolve} is
its energy determination accuracy of $\lesssim$0.2~eV at 5.9~keV \citep{Porter2025}. It
is achieved by careful design of the instrument and the operation unseen in other
high-resolution X-ray spectrometers; i.e., the $^{55}$Fe calibration sources are
illuminated intermittently during observations, and every individual event is assigned
energy based on the energy gain variation model derived for each pixel for each
observation\footnote{A report for the present observation can be found in
\url{https://heasarc.gsfc.nasa.gov/FTP/xrism/postlaunch/gainreports/3/300003010_resolve_energy_scale_report.pdf}.}.

We started with the pipeline products using processing version 03.00.011.008. For
data editing and calibration, we used \texttt{HEASoft} software build 7 and the
calibration database (CALDB) version 20230815. We restricted ourselves to using events
of the Hp grade, which are isolated in time from other events in the same pixel and thus are best
calibrated for spectroscopic purposes \citep{Porter2025,Eckart2025}. We also applied additional
screening based on the pulse height and rise time of X-ray events
\citep{Mochizuki2025a} and removed events from pixel 27, which is known to sometimes
behave anomalously in the energy gain variation. Outside the eclipse
($\phiorb=0.16$--$0.87$), the count rate in the 1.7--12~keV band is $\sim 30$~s$^{-1}$
and the Hp branching ratio is $\sim40$\%. The contributions from any contaminating source
and background events are negligible. 

\section{Analysis}\label{s3}
\subsection{Light curve and broadband spectra}\label{s3-1}
\begin{figure*}[h]
 \begin{center}
  \includegraphics[width=\textwidth]{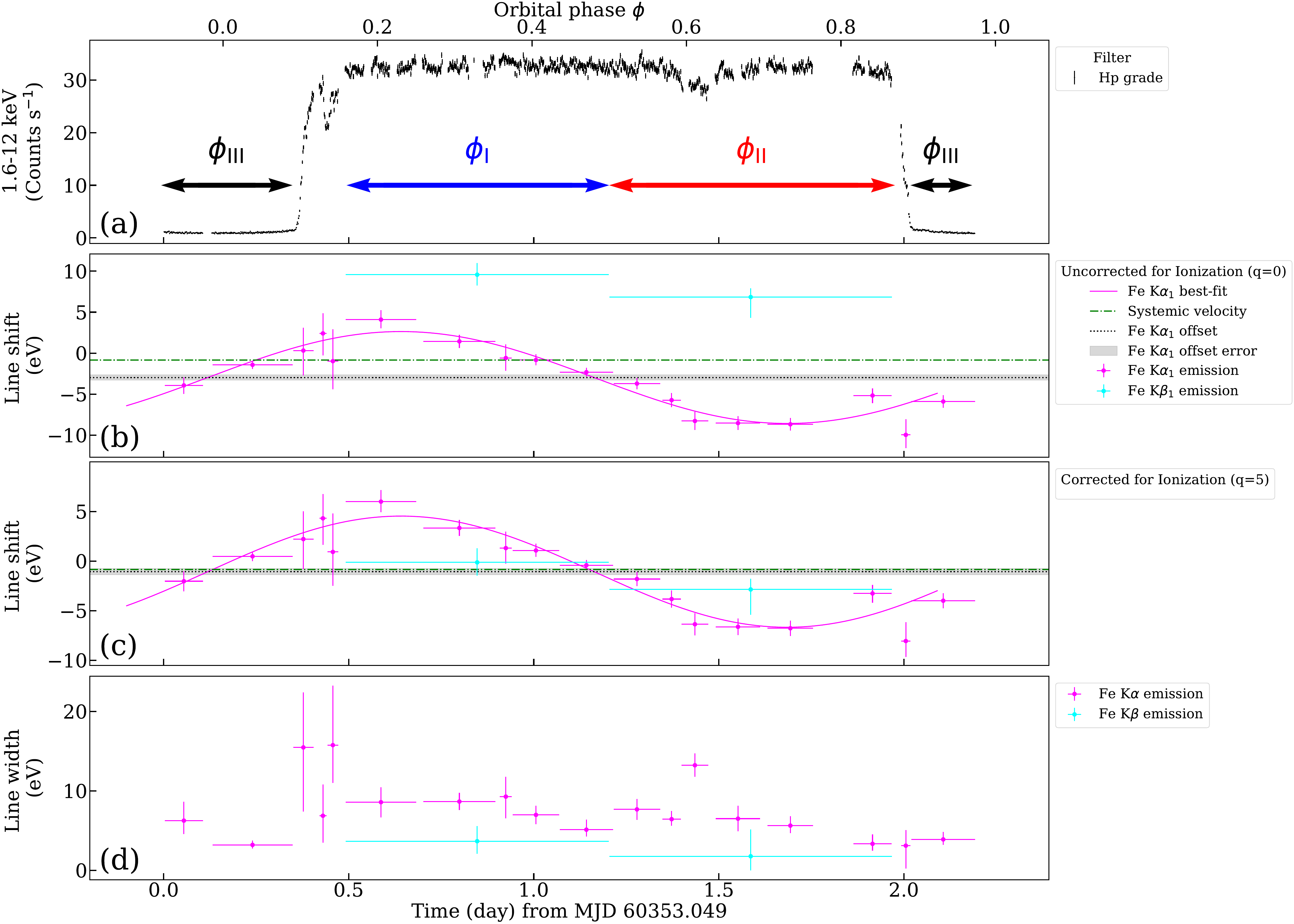}
 \end{center}
 \caption{(a) X-ray count rate, (b and c) RV curves of the Fe K$\alpha$ (magenta) and
 K$\beta$ (cyan) line before and after the ionization correction, and (d) their Gaussian
 width. The MJD is given at the bottom, while the orbital phase at the top. In panel
 (a), Hp events in the 1.6--12~keV are binned at every 128~s. The three phases
 ($\phi_\mathrm{I}$, $\phi_\mathrm{II}$, and $\phi_\mathrm{III}$) are defined with
 arrows. In panel (b) and (c), the green dashed-and-doted line is the systemic velocity
 of Cen X-3 \citep{Hutchings1979}. Panel (b) assumes that the lines are from neutral Fe
 ($q=0$), while panel (c) is corrected for the ionization effect assuming that they are
 from Sc-like Fe ($q=5$).\\ {Alt text: Four stacked line graphs with a common x-axis
 showing MJD and orbital phase. The y-axis shows count rate in the top panel, line shift
 in the middle panels, and line width in the bottom panel.}  }
 \label{fig:FeKa_orbital_variation}
\end{figure*}

Figure~\ref{fig:FeKa_orbital_variation} (a) shows the \textit{Resolve} light curves
covering the 2.09-day entire orbit of Cen~X-3. Here, the phase origin is defined at the
middle of the eclipse based on the ephemeris by \citet{Falanga2015}. We defined three
phases: $\phiorb =$0.16--0.5 ($\phi_\mathrm{I}$; 51.5~ks) when the NS is approaching us,
$\phiorb =$0.5--0.87 ($\phi_\mathrm{II}$; 48.7~ks) when the NS is retreating from us,
both outside of the eclipse, and $\phiorb=$--0.08--0.09 and 0.89--0.97
($\phi_\mathrm{III}$; 43.0~ks) in the eclipses. Some dipping behavior was observed at
$\phiorb=0.14$ and $0.61$ for $\sim 5$~ks and $\sim 9$~ks, respectively. These events
will be described in a separate paper (Loberger et al. in preparation), but are excluded
from the present analysis.

\begin{figure*}[h]
\begin{center}
 \includegraphics[width=1.02\columnwidth]{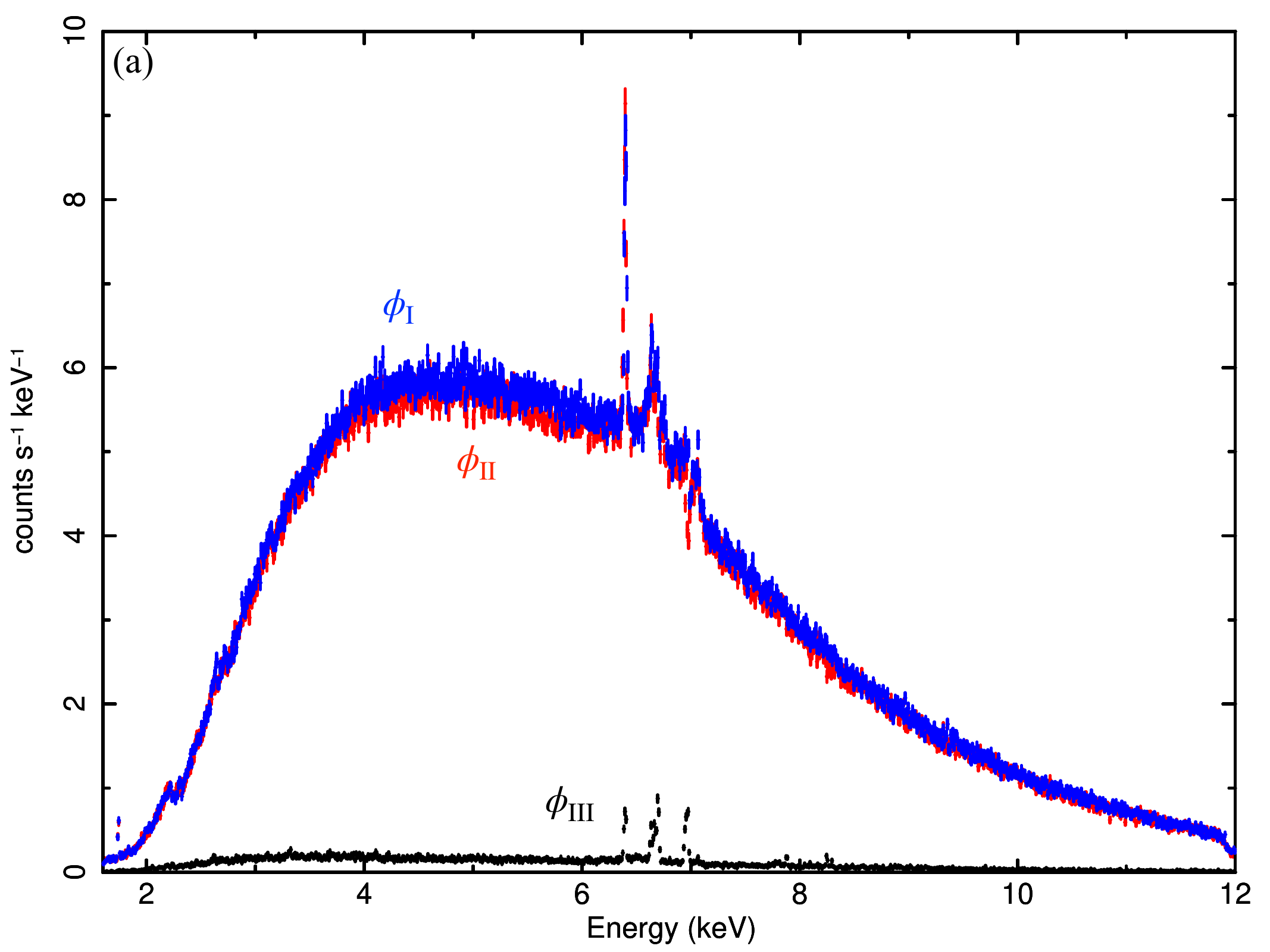}
 \includegraphics[width=0.98\columnwidth]{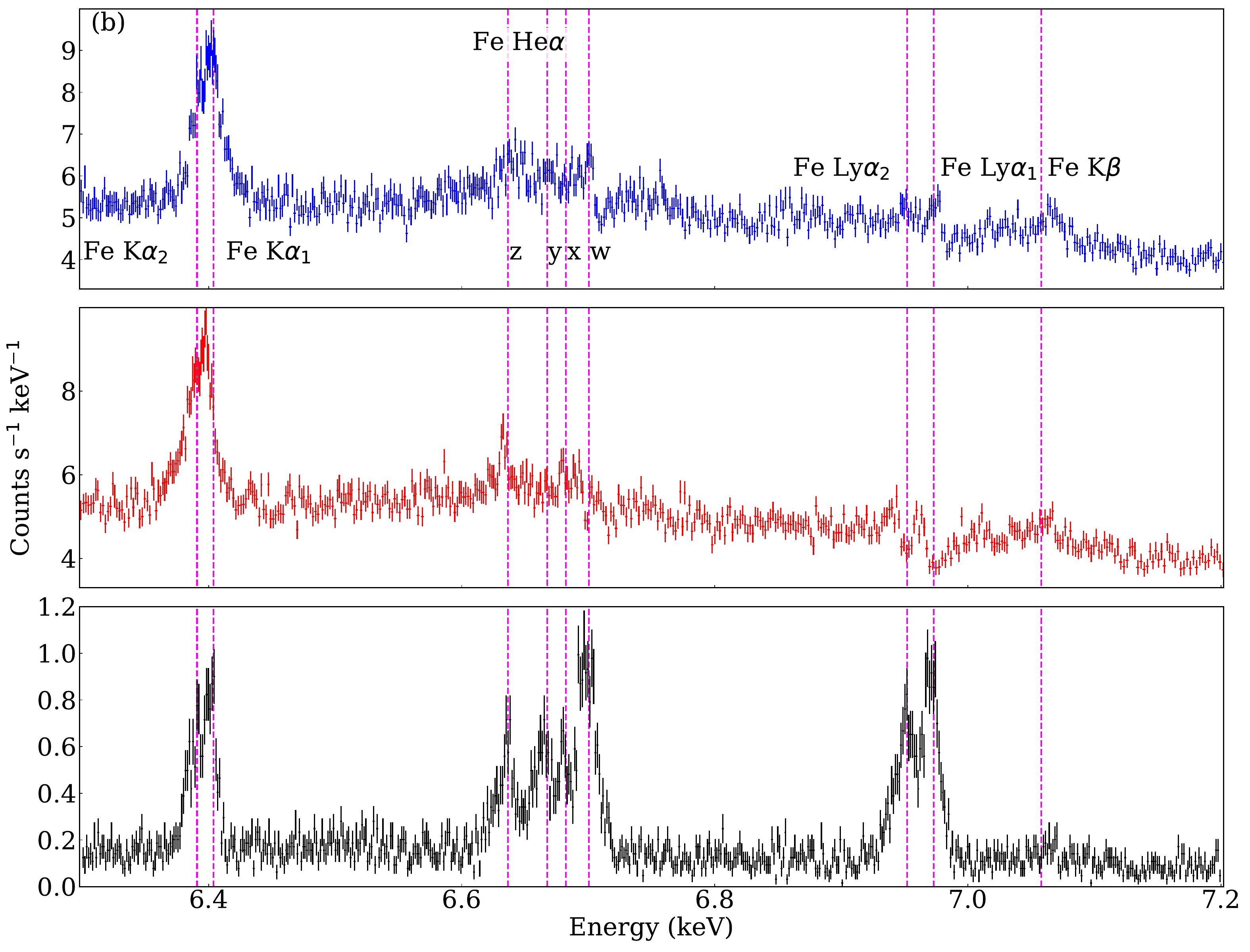}
\end{center}
 \caption{(a) Broadband spectra of \textit{Resolve} in three different phases obtained during the orbital phases  $\phi_\mathrm{I}$ (blue), $\phi_\mathrm{II}$ (red) and $\phi_\mathrm{III}$ (black). (b)
 Close-up view of each orbital-phase-resolved spectrum in the Fe K band, shown in the same color as panel a. The laboratory rest-frame energy of
 the Fe K lines are shown with the vertical magenta lines.\\ 
 {Alt text: Two line graphs show count versus energy. The right panel is divided into
 three vertical sections.}
}
\label{fig:spectrum_ironline}
\end{figure*}

The broadband spectra in each phase are shown in figure~\ref{fig:spectrum_ironline}
(a). The spectra in ($\phi_\mathrm{III}$) and out ($\phi_\mathrm{I}$ and
$\phi_\mathrm{II}$) of the eclipse are drastically different, while the two spectra in
the former ($\phi_\mathrm{I}$) and the latter ($\phi_\mathrm{II}$) half out of the eclipse
are similar except for a minor difference in the flux. We generated the detector and
mirror responses using the \texttt{rslmkrmf} and \texttt{xaarfgen} tools. In the line
spread function, the Gaussian core, its exponential tail, the Si K$\alpha$ instrumental
line, and the escape peaks were included \citep{Leutenegger2025}.

Using these responses, we fitted the out-of-eclipse spectra
($\phiorb=0.16$--$0.87$~except for $0.59$--$0.63$) in the 2.0--12~keV energy range. A
phenomenological model consisting of a power-law continuum attenuated by an interstellar
extinction described the continuum spectra reasonably well. The best-fit parameters are
the power-law photon index $1.37 \pm 0.03$, the absorption column $(3.22 \pm 0.03)
\times 10^{22}$~cm$^{-2}$, and the 2.0--12~keV band luminosity $(2.82 \pm 0.01)\times
10^{37}$~\cgslumin.

\subsection{Fe K$\alpha$ and K$\beta$ lines}\label{s3-2}
Figure \ref{fig:spectrum_ironline} (b) shows an enlarged view of the spectra around the
Fe K band. In all three phases, the Fe K$\alpha$ and K$\beta$ fluorescence lines as well
as the \ion{Fe}{XXV} He$\alpha$ and \ion{Fe}{XXVI} Ly$\alpha$ line complexes were
detected. The He$\alpha$ complex is resolved into four principal lines of $w$, $x$, $y$,
and $z$, while the Ly$\alpha$ complex is into the spin-orbit doublet of Ly$\alpha_1$ and
Ly$\alpha_2$.  Diagnostics using these lines from highly-ionized Fe are presented in
\citet{Gunasekera2025,Mochizuki2025b}. We focus on the fluorescence lines in this paper.

\begin{figure*}[h]
 \begin{center}
  \includegraphics[width=1.0\textwidth]{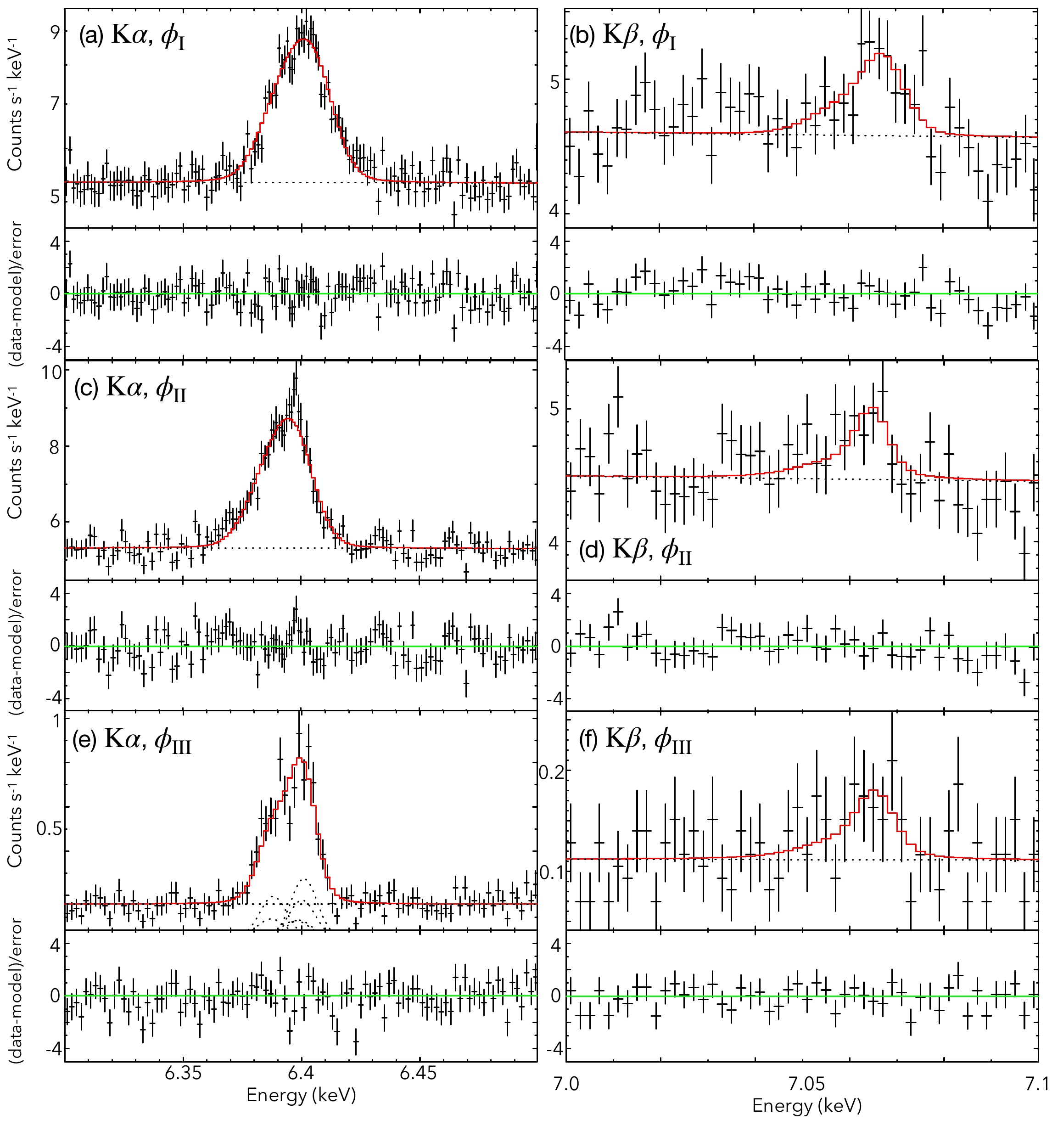}
 \end{center}
 \caption{Fitting results of the Fe K$\alpha$ (left panels) and Fe K$\beta$ (right
 panels) lines for the three phases (top: $\phi_\mathrm{I}$, middle: $\phi_\mathrm{II}$,
 and bottom: $\phi_\mathrm{III}$). In each panel, the data (black) and the best-fit
 model (red) are shown at the top, while the fitting residual at the bottom. Panels
 (a)-(b), (c)-(d), and (e)-(f) show the K$\alpha$ and K$\beta$ lines at
 $\phi_\mathrm{I}$, $\phi_\mathrm{II}$, and $\phi_\mathrm{III}$, respectively.\\
 {Alt text: Six line graphs for the K$\alpha$ and K$\beta$ lines in the three
 phases. Each panel shows the spectrum and residual to the best-fit model.}  }
 \label{fig:fitting_ironline_low}
\end{figure*}

We characterized the near-neutral Fe K$\alpha$ and K$\beta$ lines of the three phases
(figure~\ref{fig:fitting_ironline_low}), by applying the phenomenological model
\citep{Holzer1997} that is most widely used for the fluorescence line of neutral Fe. The
model consists of seven Voigt components for K$\alpha$ ($\alpha_{11}$--$\alpha_{14}$ and
$\alpha_{21}$--$\alpha_{23}$) and four for K$\beta$ ($\beta_{a}$--$\beta_{d}$). Among
them, we hereafter use the component corresponding to K$\alpha_1$ ($2p_{3/2} \rightarrow
1s$) or K$\beta_1$ ($3p_{3/2} \rightarrow 1s$)\footnote{For K$\alpha_1$, it is
K$\alpha_{11}$ in table II and figure I (c) in \citet{Holzer1997}. For K$\beta_1$, it is
K$\beta_{c}$ in their table III or K$\beta_{a}$ in figure I (d); a typo was noted in the
K$\beta$ nomenclature.} to represent the energy of each line.

We fitted the narrow-band spectra: 6.3--6.5 and 7.0--7.1 keV for the K$\alpha$ and
K$\beta$ line, respectively. For each line, we fixed the relative energy and intensity
of all components and fitted their energy shift and normalization collectively. For the
width of the Voigt profile, we fixed the Lorentz width of each component and fitted the
Gaussian width collectively for all components. To account for the underlying continuum,
we added a constant offset. This approach was successful, yielding a
statistically-acceptable fit for all three phases. There are some residuals in the line
profile, but the line centroid is tracked well enough. The best-fit parameters are
summarized in table~\ref{tab:low_ionized}.

When the energy shifts are interpreted solely due to the Doppler effect, we derived
their RV as a function of the orbital phase
(figure~\ref{fig:FeKa_orbital_variation}b). The richer statistics of the K$\alpha$ line
allowed us to conduct the spectral fitting in a finer phase binning than in three. We
obtained a sinusoidal curve for K$\alpha$ consistent with the previous work
\citep{Mochizuki2024}, including the RV offset of $-139 \pm 15$~km~s$^{-1}$. The RV
curve of K$\beta$ was presented here for the first time. The data points are too sparse
to obtain the amplitude and phase shift, but its RV offset was constrained to $345
\pm 60$~km~s$^{-1}$, which is vastly different from that of K$\alpha$.

\section{Discussion}\label{s4}
\subsection{Ionization effect of the line energy}\label{s4-1}
\begin{figure*}[h]
 \begin{center}
  \includegraphics[width=\textwidth]{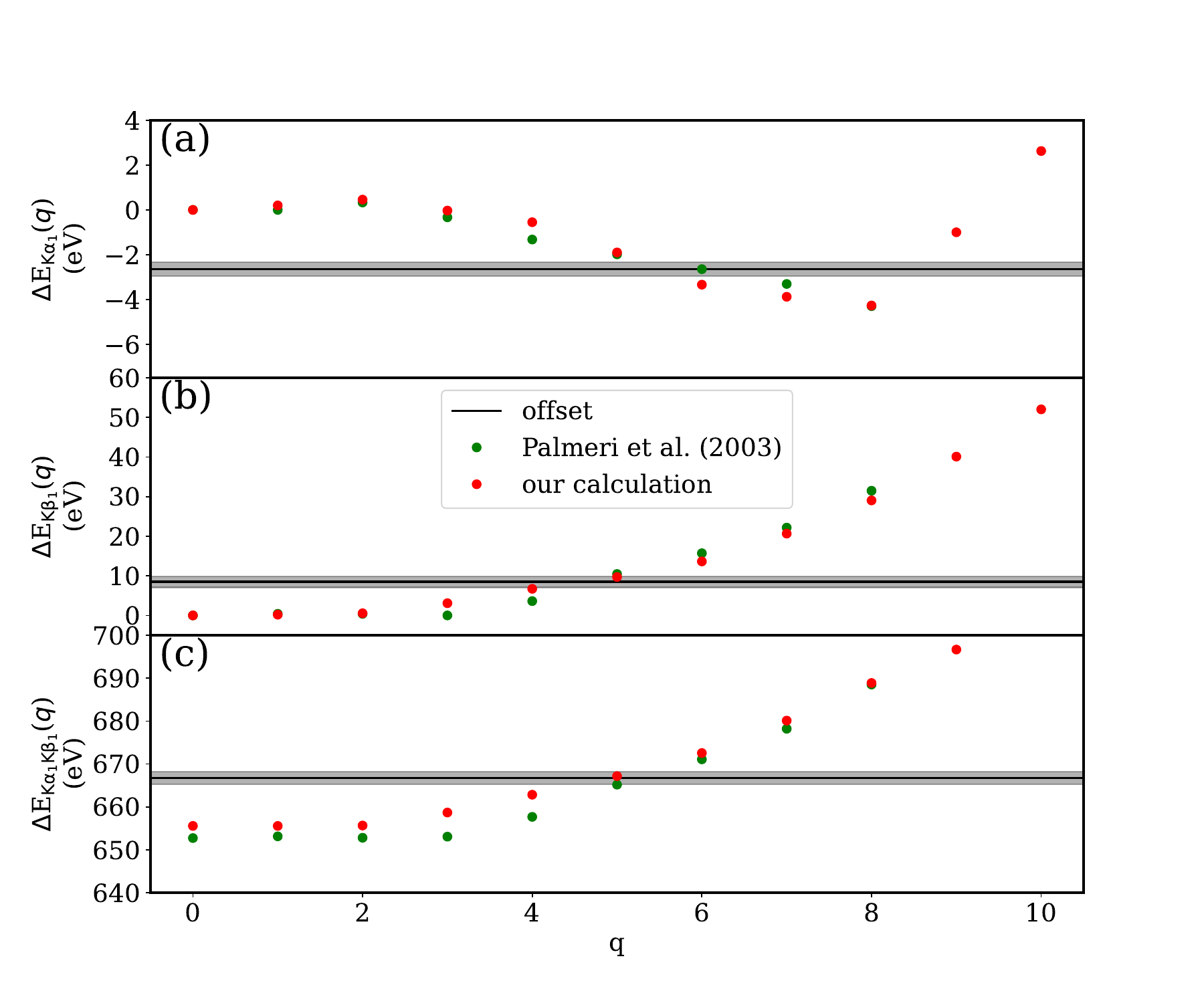}
 \end{center}
 \caption{Line energy shift of (a) K$\alpha_1$ and (b) K$\beta_1$ relative to
 the neutral value, and (c) their differences as a function of the ionization degree $q$. Two
 theoretical calculations, \citet{Palmeri2003} and this work (RLSDA/SIC), are shown in
 red and green, respectively. In the former, numerous lines in the unresolved transition array were averaged by weighting with the fluorescence yield and the results up to $q=8$ is available. In the latter calculation, a single Slater determinant was adopted without explicitly resolving angular-momentum coupling as in the configuration-interaction calculations. The resulting line energies therefore represent statistical-weighted averages over all allowed transitions, rather than those of any specific single transition. The energy difference of the initial and final orbitals are used up to $q=11$ (table~\ref{tab:rlsda_sic}).
 The horizontal black line and gray region show the observed RV offset and
 its $1\sigma$ uncertainty.\\
 {Alt text: Three line graphs with a common x-axis showing the ionization degree. The
 y-axis shows the energy difference in the top, middle, and bottom panels.}  
}
\label{fig:FeK_center_difference}
\end{figure*}

The fact that the K$\alpha$ and K$\beta$ lines exhibit different RV offsets is very
peculiar. This makes it unlikely that the shifts are dominated by any common-mode shifts
acting on both lines in the same direction, such as residual Doppler shifts. Rather, it
points toward the origins in atomic physics. In fact, such a differential-mode shift is
expected for low-ionized Fe (figure~\ref{fig:FeK_center_difference}). From $q=0$ to 2,
the energies of the K$\alpha$ and K$\beta$ lines are almost constant. From $q=3$ to 8,
the energy of the K$\alpha$ line shifts monotonically redward by $\sim$4~eV (equivalent
to $-$200~km~s$^{-1}$), while that of the K$\beta$ line shifts blueward by $\sim$30~eV
($+$1300~km~s$^{-1}$).

A simplified explanation of this effect is as follows \citep{Griffin1969}. For $q=1$ and
2, the electrons in the $4s$ orbital are removed, one by one, which has little impact on
the $3p$ and $2p$ orbitals. For $q=3$ to 8, electrons in the $3d$ orbital are removed. The
average radius of the $3d$ orbital is close to that of the $3p$ orbital.  As $q$ increases,
the energy level of the $3p$ orbital increases due to the decreasing Debye shielding of the
nuclear charge. This causes the energy of the K$\beta$ line (the transition from $3p$
to $1s$) to increase monotonically. In contrast, the average radius of the $3d$
orbital is larger than that of the $2p$ orbital. As $q$ increases, the $3d$ orbital shrinks
due to reduced repulsion among electrons in the $3d$ orbital, thus the energy level of
the $2p$ orbital decreases due to increased Debye shielding. This makes the energy of
the K$\alpha$ line (the transition from $2p$ to $1s$) decrease monotonically.

Laboratory experiments to determine the line energies of K$\alpha$ and K$\beta$ for
low-ionized Fe ($q \ge 1$) are not available in the literature. We need to rely on
atomic structure calculations, which have their own systematic uncertainties. To assess
their magnitude, we used the results of two calculations of different solvers. One is
from the literature \citep{Palmeri2003} based on the Hartree-Fock (HF) approximation
with relativistic corrections. The other is our own calculation based on the density
functional theory (DFT), or more specifically, the relativistic local-spin density
approximation with self-interaction correction (RLSDA/SIC; \cite{Perdew1981}).

The numerical procedures of the HF and DFT are similar, as both solve the
single-electron time-independent Schr\"odinger equation self-consistently. The
difference lies in how to construct the effective potential. The HF potential considers
the electron-exchange effect exactly while ignoring the electron-correlation effect
completely, whereas the DFT potential accounts for both effects using the local density
approximation \citep{Kotochigova1997}. The total energy (the energy of the entire atomic
system with a specific electron configuration, including the electron kinetic energy,
the electron–nucleus interaction, the electron–electron Coulomb repulsion, and an
approximate exchange–correlation contribution in the DFT, and the exchange-only
contribution in the HF) is consistently better in the DFT over HF calculations (table~I
in \cite{Tong1997}). The orbital energy (the energy of an electron occupying a
particular orbital within the effective potential) is also improved in the DFT over the
HF calculations \citep{Tong1998}. We present our DFT results in
table~\ref{tab:rlsda_sic}.

Still, the absolute energy of the DFT calculations is not sufficiently accurate for the
present application. The calculated values (6395.31 and 7050.87 eV for the neutral Fe
K$\alpha_1$ and K$\beta_1$, respectively) are consistently lower than the experimental
values (6403.84 and 7057.98~eV; \cite{Bearden1967}).  The discrepancy amounts to
7--8~eV. Thus, we use the differential energies to be more robust against systematic
uncertainty. Assume that $E_{\mathrm{K}\alpha_1}(q)$ and $E_{\mathrm{K}\beta_1}(q)$ are
the calculated energy of the K$\alpha_1$ and K$\beta_1$ lines of Fe$^{q+}$,
respectively. We take the difference in two ways: one is the difference against $q=0$ as
$\Delta E_{\mathrm{K}\alpha_1}(q) \equiv
E_{\mathrm{K}\alpha_1}(q)-E_{\mathrm{K}\alpha_1}(0)$ and $\Delta
E_{\mathrm{K}\beta_1}(q) \equiv E_{\mathrm{K}\beta_1}(q)-E_{\mathrm{K}\beta_1}(0)$ and
the other is the difference between the two lines of the same $q$ as $\Delta
E_{\mathrm{K}\alpha_1\beta_1}(q) \equiv
E_{\mathrm{K}\alpha_1}(q)-E_{\mathrm{K}\beta_1}(q)$. For $q=0$, $\Delta
E_{\mathrm{K}\alpha_1\beta_1}(0)$ is much more consistent between the calculation
(655.56~eV) and the experiment (654.14 eV) within 1.42~eV. For $q>1$, we have no
experimental values to compare, but the difference between the two calculations is small
enough for the present application to distinguish $q\sim0$ from $q>3$, justifying its
use. This is reasonable considering the decreased impact of the exchange and correlation
effects among electrons as $q$ increases.

Figure~\ref{fig:FeK_center_difference} shows (a) $\Delta E_{\mathrm{K}\alpha_1}(q)$, (b)
$\Delta E_{\mathrm{K}\beta_1}(q)$, and (c) $\Delta E_{\mathrm{K}\alpha_1\beta_1}(q)$ as
functions of $q$ from the two calculations.  We compared them with the RV offset
measured for the two lines and their differences. In panel (c), the measured energy
shift is most consistent with $q=5$ (Sc-like) for all three metrics, two of which are
independent from each other. The value of $\Delta E_{\mathrm{K}\alpha_1\beta_1}$ is also
consistent among the three different phases of $664.9\pm1.4$~eV ($\phi_\mathrm{I}$),
$668.9\pm1.1$~eV ($\phi_\mathrm{II}$) and $666.6\pm1.9$~eV ($\phi_\mathrm{III}$),
demonstrating that the metric is robust against varying RV shifts due to orbital motion
and can be applied even if the complete RV curve is unavailable.

In reality, we would expect some distribution in the charge state, and we consider that
$q=5$ represents the average. We use the single value for the assessments below. Taking
into account the energy shift due to the ionization effect, we revised the RV curve
modeling in figure~\ref{fig:FeKa_orbital_variation} (c) assuming $q=5$. The revised RV
offset is $-50\pm15$~km~s$^{-1}$ for K$\alpha$ and $-66\pm61$~km~s$^{-1}$ for K$\beta$,
which are now consistent with each other and also with the optical measurement of $-39$
km~s$^{-1}$ \citep{Hutchings1979}. This further reinforces our claim that the measured
line energies of K$\alpha$ and K$\beta$ are affected by the ionization effect of $q \sim
5$.

To be self-consistent, we performed the spectral fitting in subsection~\ref{s3-2} by assuming
that the lines are from Fe VI, instead of Fe I, by referring to their line properties in
table~\ref{tab:rlsda_sic}. We confirmed that, after applying this ionization correction,
the inferred average $q$ and RV offset remain the same within the statistical uncertainty.

\subsection{New constraint on the fluorescing material}\label{s4-2}
We obtained the average $q$ in the course of explaining the observed discrepancy in the
RV offset. As a byproduct, we can obtain a new constraint on the fluorescing material in
the Cen X-3 system. The inner-shell ionization source should be the NS for having
sufficient high-energy photons above the K edge energy, but the ionization source for
the fluorescent material up to $q=5$ can be different. One is the O star itself and the
other is the NS contributing also for low-ionization. 
From the amplitude of the RV curve, \citet{Mochizuki2024} argued that the location of the fluorescing material is close to the first Lagrange ($L_1$) point.
They also examined whether their model can reproduce the observed equivalent width and line width of the Fe~K$\alpha$ line as a function of the orbital phase.
In this work, we examine how the newly constrained ionization degree $q$ can contribute to further constraining such models.
To this end, we perform a simple calculation to test whether the two candidate ionization sources are both consistent with the derived value of $q$, and show that both scenarios remain plausible.

\begin{figure}[h]
 \begin{center}
  \includegraphics[width=1.0\columnwidth]{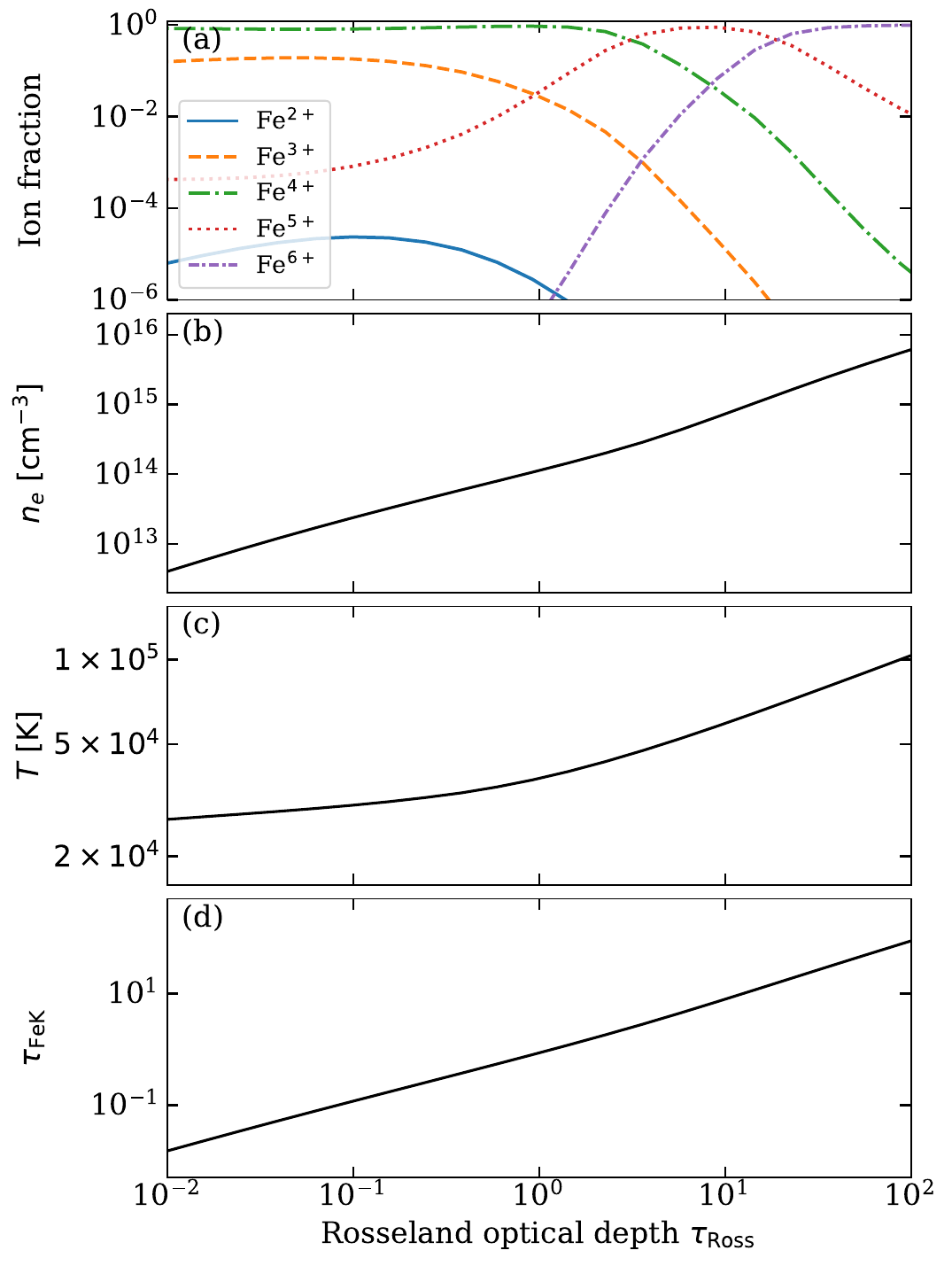}
 \end{center}
 \caption{Visualization of the atmosphere model calculation for the effective
 temperature of $3.5 \times 10^{4}$~K and the surface gravity of $10^{3.5}$ times of the
 solar value \citep{Lanz2003}: (a) the Fe charge state distribution, (b) electron
 density, (c) temperature, and (d) the Fe K optical depth ($\tau_{\mathrm{Fe K}}$) are
 shown as a function of the Rosseland optical depth ($\tau_{\mathrm{Ross}}$) as a proxy
 for the geometrical depth ($z$) from the surface. Here, $\tau_{\rm Ross}$ is the Rosseland mean optical depth, increasing inward from the stellar surface ($\tau_{\rm Ross}=0$) toward deeper atmospheric layers. The data are available online at
 \url{https://tlusty.oca.eu/tlusty/Tlusty2002/tlusty-frames-OS02.html}.
 {Alt text: Four line graphs for O star atmosphere model for the ion charge state
 distribution, electron density, temperature, and Fe K optical depth as a function of
 Rosseland optical depth.}
\label{f08}}
\end{figure}

The first is the O star. The effective temperature of 3.5$\times 10^4$~K
\citep{Hutchings1979} for a O6--8 giant appears too low (photon spectrum peaking at
5~eV) to cause ionization up to $q=5$ requiring 99~eV \citep{Kramida2024}, but because
of non-local thermodynamic equilibrium (NLTE) effects, this is possible at the
surface. \citet{Lanz2003} conducted a radiative transfer calculation using the
\texttt{TLUSTY} code for a single O star atmosphere under hydrostatic
equilibrium. Figure~\ref{f08} shows the result of the calculation, in which
(a) the Fe charge state distribution, (b) electron density, (c) temperature, and (d) the
Fe K optical depth ($\tau_{\mathrm{Fe K}}$) are shown as a function of the Rosseland
optical depth ($\tau_{\mathrm{Ross}}$) as a proxy for the geometrical depth ($z$) from the
surface. We added (d) by
\begin{equation}
  \tau_{\mathrm{Fe K}}(z) = A_{\mathrm{Fe}} \sigma_{{\mathrm{Fe K}}} \int_{0}^{z} n_{\mathrm{H}}(z)dz,
\end{equation}
in which $n_{\mathrm{H}}$ is the H number density, $A_{\mathrm{Fe}}= 6.7 \times 10^{-5}$
is the Fe abundance relative to H \citep{Wilms2000}, and $\sigma_{{\mathrm{Fe K}}}=3.8
\times 10^{-20}$~cm$^{-2}$ is the photo-electric absorption cross section at the Fe K
edge for the neutral Fe \citep{Berger2010}. The fluorescent lines are expected to be
produced most efficiently at $\tau_{\mathrm{Fe K}} \sim 1$ corresponding to
$\tau_{\mathrm{Ross}} \sim 1$ (figure~\ref{f08}d). At the depth, the charge state
distribution is mostly populated by Fe of $q=$4--6 (figure~\ref{f08}a).

\begin{figure*}[h]
 \begin{center}
  \includegraphics[width=0.9\textwidth]{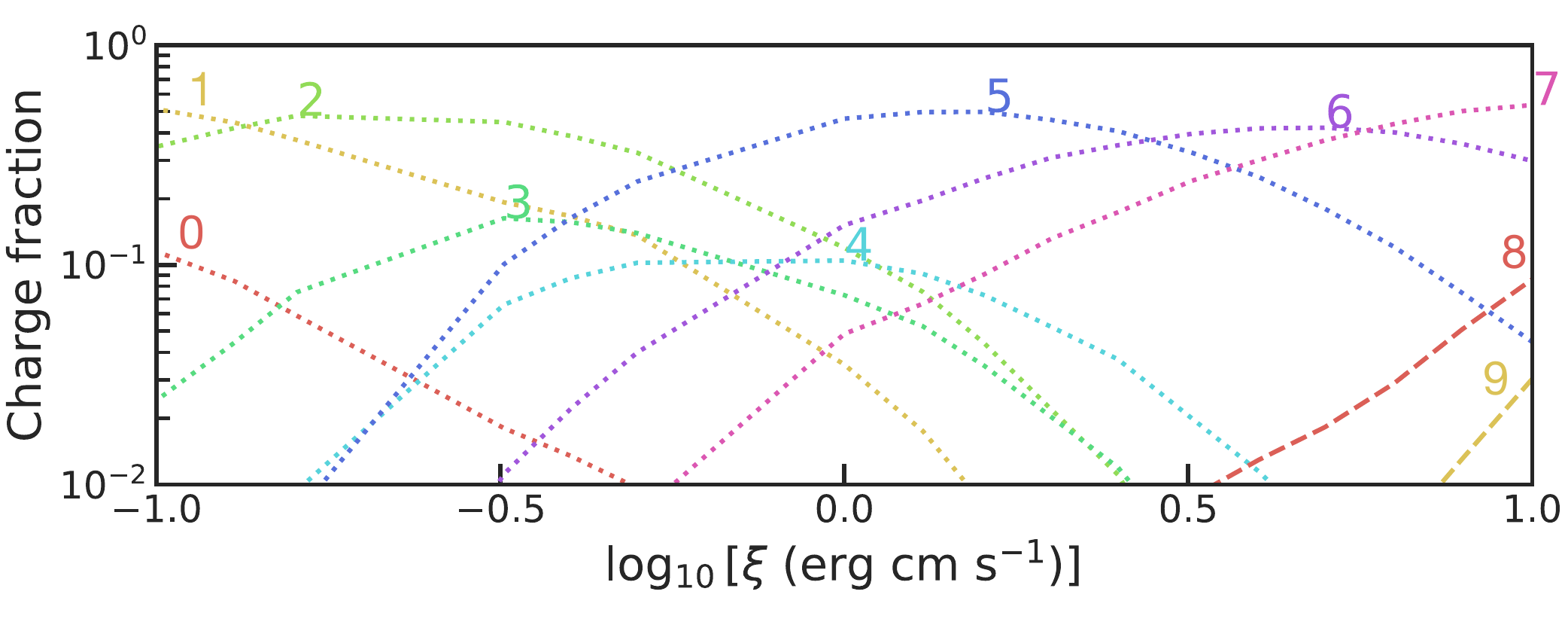}
 \end{center}
\caption{Charge state distribution (CSD) of Fe as a function of the ionization
parameter $\xi$ calculated using \texttt{XSTAR}. The number $q$ correspond to the
Fe$^{q+}$ ion. The NS spectrum of a power-law shape (photon index of 1.4) attenuated by
 a $10^{22}$~cm$^{-2}$ column representing the stellar wind is used as the photoionizing
 source.
 {Alt text: One line graph for the NS photoionization. The x-axis shows the photoionization parameter,
 and the y-axis shows the charge fraction of iron.}
\label{fig:photoionization_NS}}
\end{figure*}

The second is the NS. X-ray photons contribute to the photoionization of Fe in materials
close to $L_1$. This can be assessed using the ionization parameter \citep{Tarter1969}
defined as
\begin{eqnarray}
 \label{e01}
 \xi \equiv \frac{L}{n_er^2}~\mathrm{erg~cm~s}^{-1},
\end{eqnarray}
where $L$ (erg~s$^{-1}$) is the luminosity in 1--1000~Ryd of the ionizing source, $n_e$
(cm$^{-3}$) is the electron density of the ionized matter, and $r$ (cm) is the distance
between the ionizing source and the ionized matter. We calculated the radiative transfer
using the \texttt{XSTAR} code \citep{Kallman2001} to obtain the charge state
distribution of Fe as a function of $\xi$ (figure~\ref{fig:photoionization_NS}). In the
calculation, we assumed that the spectral shape of the NS is represented by a power-law
of a photon index of 1.4 (subsection~\ref{s3-1}) and the incident emission went through the
stellar wind represented by a slab of a column having $10^{22}$~cm$^{-2}$ before
reaching the O star surface. Note that the column is sufficiently thin
($\tau_{\mathrm{FeK}} \sim 10^{-2}$) for the inner-shell ionizing photons. The maximum
formation of Fe at $q=5$ is achieved around $\log_{10}{\xi} = 0$. For $L=3.2 \times
10^{37}$~erg~s$^{-1}$ from the observation (subsection~\ref{s3-1}) and $r=4.4 R_{\odot}$ for the
distance from the NS to the $L_1$ point, this converts to $n_{\mathrm{e}} \sim 3 \times
10^{14}$~cm$^{-3}$. This density is achieved again at $\tau_{\mathrm{Ross}} \sim 1$
(figure~\ref{f08}b), suggesting that photoionization by NS can also contribute for the
ionization to $q \sim 5$.

\section{Conclusions and Prospects}\label{s5}
The Fe K$\alpha$ fluorescence line at 6.4~keV is widely used in X-ray spectroscopy. The
diagnostic is used for the X-ray microcalorimeter data in ways that were never possible
before. We need to know the intrinsic properties of the line to the accuracy that the
XRISM data can achieve. One caution was found in the RV curve of Cen X-3, in which the
RV offset of K$\alpha$ is significantly larger than the optical measurement and is
inconsistent with that of K$\beta$. We revisited the assumption that these lines are
produced by neutral ($q=0$) Fe. As the ionization increases, the K$\alpha$ line shifts
redward while the K$\beta$ line shifts blueward. This differential shift is distinctive
to the ionization effect, and we used the energy difference as a metric to constrain the
most likely charge state. By applying the method, we constrained $q \sim 5$. By
correcting the line shift of the K$\alpha$ and K$\beta$ lines, we solved the discrepancy
in the RV offset of Cen X-3. This demonstrates that (1) the ionization effect is
actually observable in microcalorimeter data and (2) the differential shift between the
K$\alpha$ and K$\beta$ lines serves as a robust metric of the effect.

The ionization effect should occur and should be assessed in all use cases of Fe
K$\alpha$ diagnostics beyond Cen X-3. An example is the immediate post-shock plasmas in
supernova remnants. One can constrain both the bulk motion and the charge state of the
shocked plasma, thereby providing valuable insights into the efficiency of collisionless
electron heating (e.g., \cite{Yamaguchi2014}). Another example is the Fe K$\alpha$ line
from the surfaces of compact objects such as white dwarfs. One can constrain the
gravitational redshift, which thereby offers a means to constrain the mass of compact
objects (e.g., \cite{Hayashi2023}). The proposed metric of the differential shift
between the K$\alpha$ and K$\beta$ lines will be generically useful in all these cases.

\begin{ack}
 We express our gratitude to all scientists and engineers for their persistent efforts
 to make XRISM possible, following the previous X-ray microcalorimeter missions onboard
 Astro-E, Suzaku, and Hitomi in international collaborations. 
 Y.\,N. is supported by the JST SPRING program (grant number JPMJSP2110), 
 T.\,E. by the JST Sohatsu program (grant number JPMJFR202O), 
 H.\,Y. by JSPS KAKENHI (grant number 22H00158), 
 Y.\,M. by the JST SPRING program (grant number JPMJSP2108) and JSPS KAKENHI (grant number JP25KJ0923),
 E.\,B. by the Israel Science Foundation (grant number 2617/25),
 and L.\,C. by NASA (grant number 80NSSC18K0978, 80NSSC20K0883, and 80NSSC25K7064). N.\,H.'s work was performed under the auspices of the U.S. Department of Energy by Lawrence Livermore National Laboratory under Contract DE-AC52-07NA27344.
\end{ack}

\clearpage 
\begin{table*}
\tbl{Summary of the system parameters of Cen~X-3 and the XRISM observation.}
{
\begin{tabular}{p{20em}cll}
\hline \hline
Parameter & Symbol & Value & Reference \\ 
\hline
\multicolumn{4}{l}{Binary system parameters} \\ 
Mass of the primary star \dotfill & $M_{*}$ & $20.5\pm 0.7M_{\odot}$ & \citet{Ash1999} \\
Radius of the primary star \dotfill & $R_{*}$ & $11.8R_{\odot}=8.22\times 10^{11}$~cm  & \citet{Wojdowski2001}\\
Spectral type of the primary star \dotfill & -- & O6--O8 (f)  & \citet{Hutchings1979}\\
Surface temperature of the primary star \dotfill & $T_{*}$ & $35000\pm2000$~K  & \citet{Hutchings1979}\\
Bolometric luminosity of the primary star \dotfill & $L_{*}$ & $\sim1.9\times 10^5 L_{\odot}=7.2\times 10^{38}$~erg s$^{-1}$ & derived using the Stefan–Boltzmann law\\
Mass of the NS \dotfill & $M_{\textrm{x}}$ & 1.21$\pm$ 0.21$M_{\odot}$ & \citet{Ash1999}\\
Spin period of the NS\footnotemark[$*$] \dotfill & $P_{\textrm{s}}$ & 4.791852~s & this work \\
Orbital period \dotfill & $P_{\textrm{orb}}$ & $2.08704106\pm0.00000003$~days & \citet{Falanga2015}\\
Binary separation \dotfill & $a$ & $18.14R_\odot$ & \citet{Bildsten1997}\\
Lagrange point $L_1$ distance from the NS \dotfill & -- & $\sim$ 4.4$R_\odot$ & derived using \citet{Eggleton1983} \\
Inclination angle \dotfill & $i$ & $(70.2\pm2.7)$~degree & \citet{Ash1999} \\
Distance \dotfill & $d$ & 6.4~kpc & \citet{Tsygankov2022}\\
Systemic velocity\footnotemark[$**$] \dotfill & $\gamma$ & $-39$~km~s$^{-1}$ & \citet{Hutchings1979} \\
Position (equinox J2000.0) \dotfill & (RA, Dec) & (170.3128$^{\circ}$, $-$60.6237$^{\circ}$) \\ 
\hline
\multicolumn{4}{l}{Observation summary of XRISM} \\ 
ObsID \dotfill & & \multicolumn{2}{l}{300003010} \\
Start and end of observations \dotfill & & \multicolumn{2}{l}{2024-02-12 23:56:04 to 2024-02-15 06:19:04 UT} \\
Observation duration in orbital phase \dotfill & $\phi_{\textrm{orb}}$ & $-0.08$ to $0.97$ \\ 
Total exposure time \dotfill & & 196~ks\\
2.0--12~keV X-ray flux\footnotemark[$\dag$] \dotfill & $F_{\textrm{x}}$ & $ (4.858\pm0.004)\times10^{-9}$~\cgsflux \\ 
2.0--12~keV X-ray luminosity\footnotemark[$\dag$]
 \dotfill & $L_{\textrm{x}}$ & $ (2.82\pm0.01)\times 10^{37}$~\cgslumin \\ 
\hline
\multicolumn{4}{l}{Radial velocity (RV) curve with K$\alpha$} \\ 
RV amplitude \dotfill & & $263 \pm 20$ km s$^{-1}$ ($5.60\pm 0.43$ eV in energy)& this work \\
RV offset ($q=0$)\footnotemark[$\ddag$] \dotfill & & $-139\pm 15$ km s$^{-1}$ ($-3.0\pm 0.3$ eV in energy)& this work\\
RV offset ($q=5$)\footnotemark[$\S$] \dotfill & & $-50\pm 15$ km s$^{-1}$ ($-1.1\pm 0.3$ eV in energy)& this work \\
\hline
\end{tabular}
}

\label{tab:par_summary}
\begin{tabnote}
\footnotemark[$*$] The spin period at MJD 60353. \\
\footnotemark[$**$] The binary system is receding from us. \\
\footnotemark[$\dag$] The absorbed flux and unabsorbed luminosity are the values in the
 2--12~keV band during the out-of-eclipse except dip phases.  \\ 
\footnotemark[$\ddag$] Assuming that the line is emitted from neutral Fe ($q=0$).\\
\footnotemark[$\S$] Assuming that the line is emitted from Sc-like Fe ($q=5$).\\
\end{tabnote}
\end{table*}

\begin{table*}
\tbl{Best-fit parameters of the Fe K$\alpha_{11}$ and K$\beta_{c}$ lines using the
 \citet{Holzer1997} model.}
{
\begin{tabular}{lcccccc}
\hline \hline
Line\footnotemark[$\dag$] & Ref. Energy\footnotemark[$\dag$] & Det. Energy\footnotemark[$\ddag$] & Velocity\footnotemark[$\S$] & Width & Flux\footnotemark[$*$] & EW\footnotemark[$*$] \\ 
& (eV) & (eV) & (km~s$^{-1}$) & (eV) & (ph~s$^{-1}$~cm$^{-2}$) & (eV) \\
\hline
\multicolumn{7}{l}{$\phi_\mathrm{I}=0.16$--$0.5$} \\ 
Fe K$\alpha_{11}$  & $6404.15$ & $6404.7^{+0.4}_{-0.4}$ & $ (+2.6^{+1.8}_{-1.8})\times10^{1}$ & $8.4^{+0.6}_{-0.6}$ & $ (2.9^{+0.1}_{-0.1})\times10^{-4}$ & $4.8^{+0.2}_{-0.1}$ \\
Fe K$\beta_{c}$ & 7058.36 & $7068.3^{+1.3}_{-1.4}$ & $ (+4.2^{+0.6}_{-0.6})\times10^2$ & $4^{+2}_{-2}$ & $ (2.8^{+0.6}_{-0.6})\times10^{-5}$
& $0.5^{+0.1}_{-0.1}$ \\
\hline
\multicolumn{7}{l}{$\phi_\mathrm{II}=0.5$--$0.87$} \\ 
Fe K$\alpha_{11}$ & 6404.15 & $6398.1^{+0.4}_{-0.4}$ & $ (-2.9^{+0.2}_{-0.2})\times10^2$ & $7.6^{+0.6}_{-0.6}$ & $ (2.8^{+0.1}_{-0.1})\times10^{-4}$ & $4.8^{+0.2}_{-0.2}$ \\
Fe K$\beta_{c}$ & 7058.36 & $7065.5^{+2.5}_{-1.1}$ & $ (+3.0^{+0.5}_{-1.1})\times10^2$ & $<5.2$ & $ (2.1^{+0.7}_{-0.5})\times10^{-5}$ & $0.4^{+0.1}_{-0.1}$ \\
\hline 
\multicolumn{7}{l}{$\phi_\mathrm{III}=-0.08$--$0.09$, $0.89$--$0.97$} \\ 
Fe K$\alpha_{11}$ & 6404.15 & $6401.0^{+0.5}_{-0.4}$ & $ (-1.5^{+0.2}_{-0.2})\times10^2$ & $4.9^{+0.5}_{-0.5}$ & $ (2.4^{+0.1}_{-0.1})\times10^{-5}$ & $23^{+2}_{-2}$ \\
Fe K$\beta_{c}$ & 7058.36 & $7066.3^{+2.1}_{-1.8}$ & $ (+3.4^{+0.8}_{-0.9})\times10^2$ & $<5.3$ & $ (1.6^{+0.6}_{-0.5})\times10^{-6}$ & $2.4^{+0.8}_{-0.9}$ \\
\hline
\end{tabular}
}
\label{tab:low_ionized}
\begin{tabnote}
\footnotemark[$\dag$] Reference energy \citep{Holzer1997}. In this referenced model, the Fe K$\alpha$ and Fe K$\beta$ lines are approximated by seven and four Voigt functions, respectively. Among these components, the strongest ones are the Fe K$\alpha_{11}$ and Fe K$\beta_{c}$ lines, respectively. \\
\footnotemark[$\ddag$] Observed energy derived by the fitting.\\
\footnotemark[$\S$] The velocity when the energy shift is interpreted solely due to the Doppler shift. \\
\footnotemark[$*$] Values of K$\alpha_{11}$ and K$\beta_{c}$ only, not the entire complex. \\  
\end{tabnote}
\end{table*}

\clearpage 

\begin{table*}
\tbl{Fe K line\footnotemark[$\dag$]  energies and Einstein A coefficients calculated with the RLSDA/SIC method.}
{
 \begin{tabular}{ccccccccc}
 \hline
 \hline
 $q$ &
 $E_{\mathrm{K}\alpha_1}$ &
 $E_{\mathrm{K}\alpha_2}$ &
 $E_{\mathrm{K}\beta_1}$ &
 $E_{\mathrm{K}\beta_3}$ &
 $A_{\mathrm{K}\alpha_1}$ &
 $A_{\mathrm{K}\alpha_2}$ &
 $A_{\mathrm{K}\beta_1}$ &
 $A_{\mathrm{K}\beta_3}$ \\
 &
 (eV) &
 (eV) &
 (eV) &
 (eV) &
 (s$^{-1}$) &
 (s$^{-1}$) &
 (s$^{-1}$) &
 (s$^{-1}$) \\
 \hline
 0  & 6395.31 & 6382.37 & 7050.87 & 7049.27 &
 $1.70721\times10^{14}$ & $1.74333\times10^{14}$ &
 $2.04859\times10^{13}$ & $2.08614\times10^{13}$ \\
 1  & 6395.51 & 6382.57 & 7051.06 & 7049.47 &
 $1.70741\times10^{14}$ & $1.74354\times10^{14}$ &
 $2.04545\times10^{13}$ & $2.08293\times10^{13}$ \\
 2  & 6395.77 & 6382.83 & 7051.42 & 7049.82 &
 $1.70773\times10^{14}$ & $1.74386\times10^{14}$ &
 $2.04297\times10^{13}$ & $2.08026\times10^{13}$ \\
 3  & 6395.28 & 6382.34 & 7053.96 & 7052.31 &
 $1.70709\times10^{14}$ & $1.74321\times10^{14}$ &
 $2.10084\times10^{13}$ & $2.13679\times10^{13}$ \\
 4  & 6394.76 & 6381.81 & 7057.59 & 7055.87 &
 $1.70645\times10^{14}$ & $1.74253\times10^{14}$ &
 $2.18021\times10^{13}$ & $2.21472\times10^{13}$ \\
 5  & 6393.41 & 6380.48 & 7060.58 & 7058.77 &
 $1.70595\times10^{14}$ & $1.74199\times10^{14}$ &
 $2.27782\times10^{13}$ & $2.31095\times10^{13}$ \\
 6  & 6391.97 & 6379.05 & 7064.51 & 7062.60 &
 $1.70572\times10^{14}$ & $1.74169\times10^{14}$ &
 $2.39024\times10^{13}$ & $2.42208\times10^{13}$ \\
 7  & 6391.43 & 6378.49 & 7071.53 & 7069.53 &
 $1.70584\times10^{14}$ & $1.74174\times10^{14}$ &
 $2.51505\times10^{13}$ & $2.54572\times10^{13}$ \\
 8  & 6391.04 & 6378.08 & 7079.91 & 7077.81 &
 $1.70649\times10^{14}$ & $1.74229\times10^{14}$ &
 $2.65037\times10^{13}$ & $2.67998\times10^{13}$ \\
 9  & 6394.31 & 6381.29 & 7090.97 & 7088.76 &
 $1.71267\times10^{14}$ & $1.74844\times10^{14}$ &
 $2.78286\times10^{13}$ & $2.81127\times10^{13}$ \\
 10 & 6397.94 & 6384.87 & 7102.90 & 7100.58 &
 $1.71966\times10^{14}$ & $1.75539\times10^{14}$ &
 $2.91842\times10^{13}$ & $2.94573\times10^{13}$ \\
 11 & 6402.03 & 6388.89 & 7117.89 & 7115.47 &
 $1.72748\times10^{14}$ & $1.76317\times10^{14}$ &
 $3.05707\times10^{13}$ & $3.08335\times10^{13}$ \\
 \hline
 \end{tabular}}
\label{tab:rlsda_sic}
\begin{tabnote}
 \footnotemark[$\dag$] The transition of the lines are K$\alpha_1$ ($2p_{3/2}
 \rightarrow 1s$), K$\alpha_2$  ($2p_{1/2} \rightarrow 1s$), K$\beta_1$  ($3p_{3/2}
 \rightarrow 1s$), and K$\beta_3$ ($3p_{1/2} \rightarrow 1s$). The nomenclatures follow
 \citet{Holzer1997,Palmeri2003}. 
\\
\end{tabnote}
\end{table*}


\clearpage 

\clearpage 


\clearpage 

\clearpage 

\clearpage 

\clearpage 

\end{document}